\documentclass[article]{elsarticle}

\usepackage{lineno,hyperref}
\modulolinenumbers[5]

\journal{ArXiv}

\usepackage{chemformula}
\usepackage{siunitx}
\usepackage{caption}

\usepackage{float}
\usepackage[hang]{subfigure}
\usepackage{todonotes}

\begin{document}

\begin{frontmatter}

\title{Electronic band structure screening for Dirac points in Heuslers}

\author[1]{Meza-Morales, Paul J.}
\author[1]{Fumarola, Alessandro}
\author[1]{Taliaronak, Volha}
\author[1]{Shirsekar, Afrid}
\author[1]{Kidner, Jonathan}
\author[1]{Ali, Zaheer}
\author[1]{Ali, Mazhar\corref{cor1}}
\address[1]{Material Mind, Fremont, CA 94555, United States}
\cortext[cor1]{Corresponding author}

\begin{abstract}
The  Heusler compounds have provided a playground of material candidates for various technological applications based on their highly diverse and tunable properties, controlled by chemical composition and crystal structure. However, physical exploration of the Heusler chemical space \textit{en masse} is impossible in practice, hindering the exploration of the chemical composition vs. proprieties relationship. Many of these applications are related to the Heuslers electron transport characteristics, which are embedded in their electronic band structure (EBS). Here we we created a Heuslers dataset using the Materials Project (MP) database --- retrieving both chemical composition and their EBSs. We then used machine learning to develop a model correlating the composition vs. number of Dirac points in the EBS for Heuslers and also other Cubic compounds by identifying said Dirac points using an automated algorithm as well as generating chemical composition and global crystal structure features. Our ML model captures the overall trend, as well as identifies significant electronic and global crystal structure features, however, the ML model suffered from  significant variance due to the lack of site specific features. Future work on a methodology for handling atomic site specific features will allow ML models to better match the underlying quantum mechanics governing the properties (also based on site specific properties) and capture the electronic properties in a more generalized approach.

\end{abstract}


\end{frontmatter}

\nolinenumbers

\section{Introduction}
\begin{figure}[H]
    \centering
        \subfigure[]{
        \includegraphics[width=3.0in]{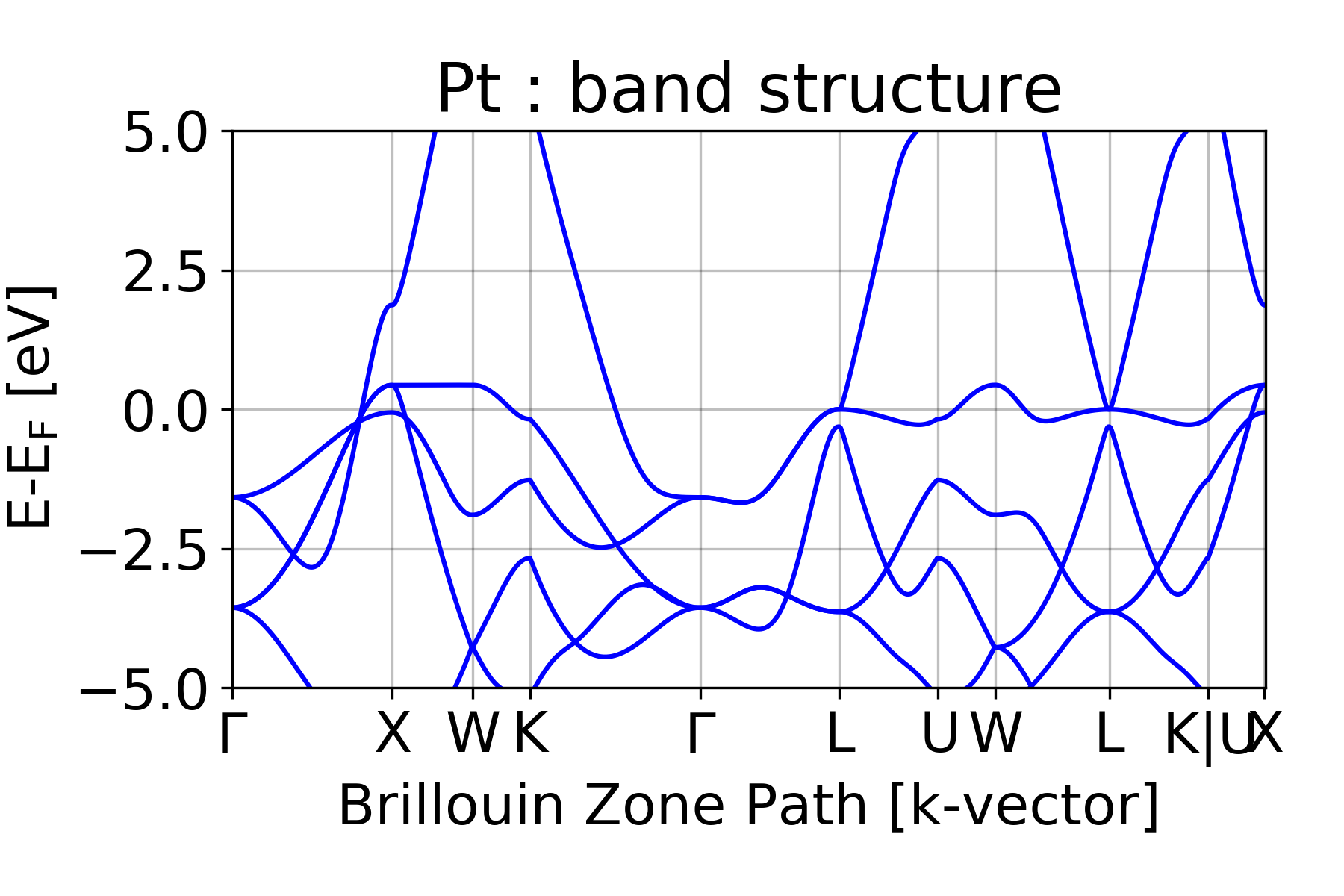}
        \label{fig:Pt_EBS}
        }
        
        \subfigure[]{
        \includegraphics[width=3.0in]{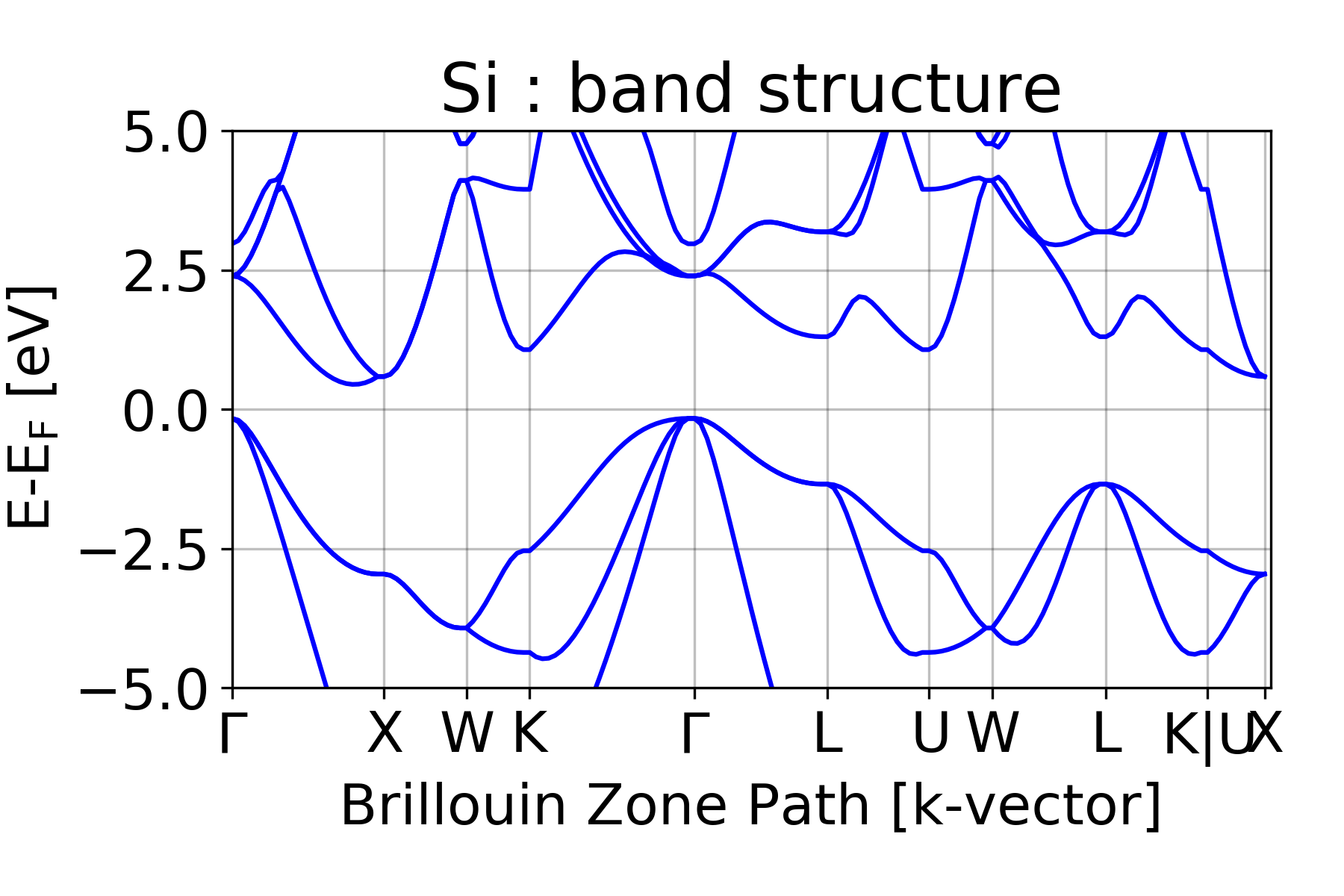}
        \label{fig:Si_EBS}
        }
        
        \subfigure[]{
        \includegraphics[width=3.0in]{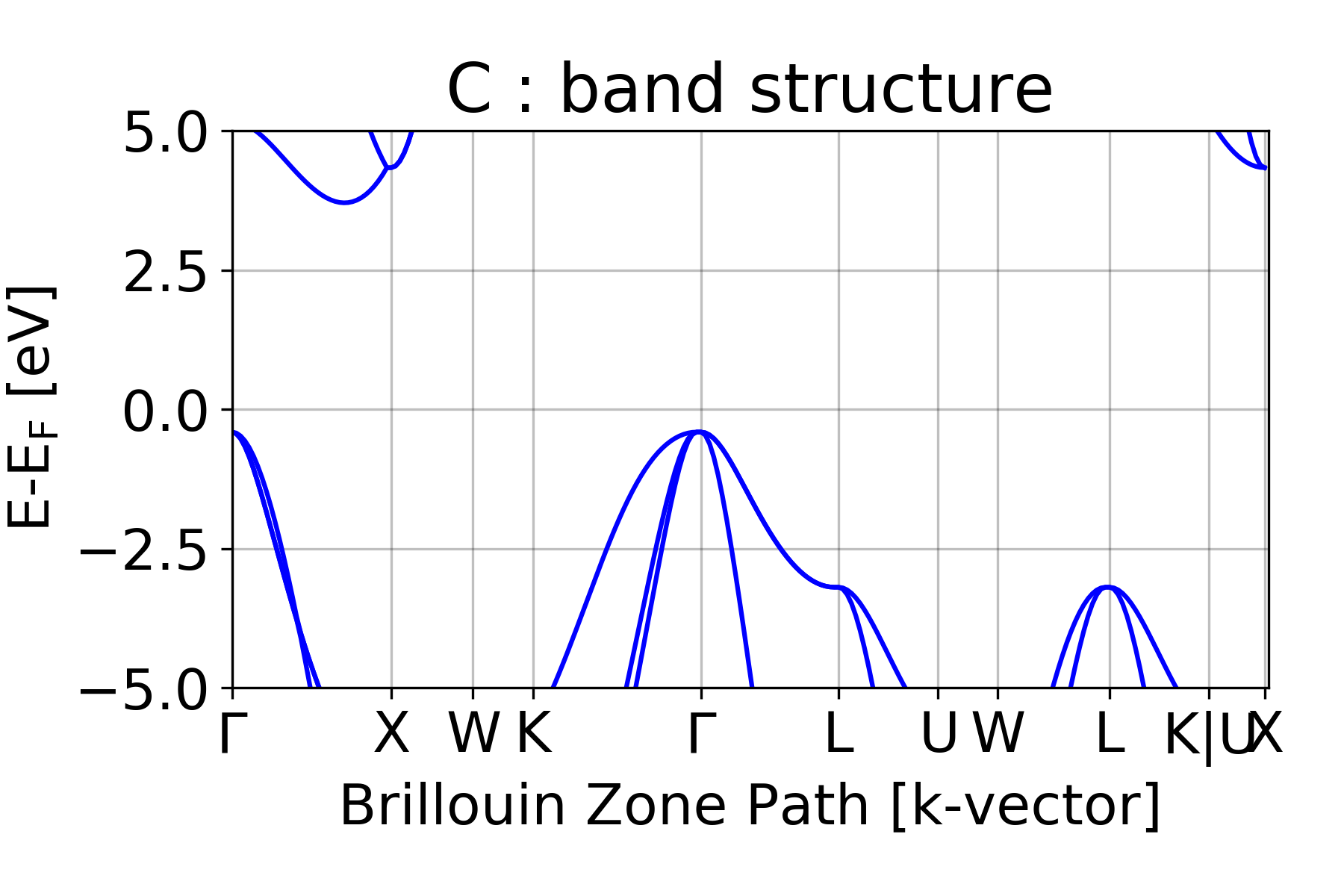}
        \label{fig:Diamond_EBS}
        }
    \caption{comparison of EBSs for metallic vs semiconducting vs insulating behavior using: a) metallic (\ch{Pt}), b) semiconducting (\ch{Si}), and c) insulating (Diamond). The energies of the band gaps are \SI{4.34}{\electronvolt} for Diamond,  \SI{0.85}{\electronvolt} for \ch{Si} and \SI{0.0}{\electronvolt} for \ch{Pt}. These EBSs were taken from Materials Project (MP) database \cite{Jain2013,jain2013commentary,ong2015materials} their MP-id are included in the legend of each material. [colors online]}
    \label{fig:EBS_classical}
\end{figure}

\begin{figure}[H]
    \centering
    \includegraphics[width=0.7\textwidth]{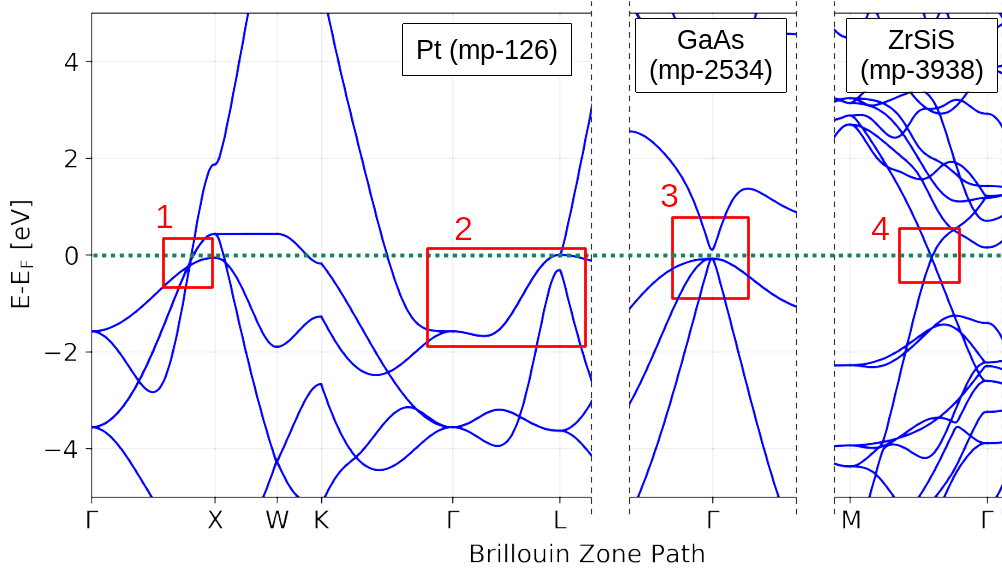}
    \caption{Deeper patterns in the EBS for Platinum (\ch{Pt}), Gallium arsenide (\ch{GaAs}), and Zirconium Silicon Sulfide (\ch{ZrSiS}). Red boxes correspond to 1. tilted Dirac point, 2. Nodal line, 3. local flat band with other highly dispersive bands, and 4. Dirac point at the Fermi level. The EBSs were taken from Materials Project (MP) database,\cite{Jain2013} their MP-id are included in the legend of each material. [colors online]}
    \label{fig:EBS_deeper_patterns}
\end{figure}

Quantum materials (QMs)---materials in which quantum effects manifest non-classical properties---are expected to drive a variety of next generation technologies ranging from quantum computing\cite{bassman2021simulating,han2018quantum,lau2020emergent} and sensing\cite{zhu2019progress,hennighausen2021twistronics,crawford2021quantum} to novel energy storage,\cite{broholm2016basic,tokura2017emergent,koo2019research,lau2020emergent} telecommunication\cite{birsan2020zr,wei2019machine,khandy2019full,grigaliunaite2020magnetic,fowley2018magnetocrystalline} and renewable energy generation solutions\cite{kamlesh2021first,kamlesh2022comprehensive,zaferani2019strategies}. The electronic properties of these QMs are derived from their electronic band structures (EBSs); the spectrum of electronic energy states vs electron momenta, arising from the solutions of Schrodinger's equation for electrons in a periodic potential (i.e. crystal lattice, a periodic arrangement of atoms in space). For example, a classical material property, like metallic vs semiconducting vs insulating behavior, (see Figure \ref{fig:EBS_classical}) can be quickly understood by observing the energetic ``band gaps'' in the EBS. However in QMs, various non-classical properties can be gleaned from deeper patterns in the EBS such as ``flat bands'', ``Dirac point'', ``Weyl points'', ``Nodal Lines'', or spin-orbit coupled gap openings. For example, ``flat bands" (Figure \ref{fig:EBS_deeper_patterns}) are known to be associated with strong electron correlation and can contribute to properties like superconductivity, charge density wave ordering, magnetism, and more\cite{balents2020superconductivity,yang2021testing,hu2021charge}. On the other hand, highly dispersive linear bands intersecting at a crossing point (a.k.a. Dirac point, Figure \ref{fig:EBS_deeper_patterns}), are known to be associated with high carrier mobility, large Hall effects, spatially localized conduction and more\cite{falson2018review,zhang2021cycling,polash2021topological,Liang2015}. Combinations of Dirac points, Weyl points, Nodal Lines, flat band, spin-orbit coupled gap openings, and other patterns give rise to higher order electronic properties that are relevant to next generation materials and technologies. Hence predicting, identifying and understanding the patterns in the EBS is of great interest to the condensed matter, device physics and communities.

Currently, electronic band structures are calculated using Density Functional Theory (DFT), a powerful method for quantum mechanical calculations of many body systems. While successful at generating accurate results for nonmagnetic and low spin-orbit coupled systems, DFT calculations become computationally expensive as non-idealities of the model are considered. This includes crystals defects, magnetism, strong spin-orbit coupling (SOC), large unit cells, strong electron correlation (J), etc. It is also time consuming to carry out these calculations en-masse; while the Materials Project\cite{ward2018matminer} has approximately 100K materials in the database, this is a small number compared to the more than \num{d18} possible material combinations predicted to exist, and carrying out full DFT + SOC + U + magnetism calculations on that dataset is simply unrealistic. While supercomputers and new exchange-correlation functional development aid in expanding the reach of DFT, another compelling approach may be to use artificial intelligence (AI) algorithms---such as machine learning (ML)---to predict patterns in EBSs from a chemical composition and crystal structure without carrying out intensive DFT calculations. Thus, by training on known DFT computed EBSs and learning what chemical compositions/structural features drive certain electronic structure patterns, it is possible to use AI models as a rapid screening tool and identify promising material classes or candidates for deeper theoretical and experimental investigation as well as to discover new materials with desired properties. These machine learning methods center around the core idea of learning against the hidden correlations of the material's structures with the physical property of relevance for a subgroup of materials, where the property values are either computationally computed or experimentally available, to quantitatively establish a structure-property relationship. The insight gained from the subgroup of materials is then utilized to deliver predictions on the property for all materials in the parent group, thus eliminating the need for spending time consuming computational or experimental resources for these structures.\cite{moghadam2019structure,sun2019machine,zhou2019big,goldsmith2018machine,toyao2019machine} In the case where there is a paucity of experimental physical property data but a particular pattern in the electronic band structure is known to relate to a physical property of interest, the correlation methodology can become focused on learning against the correlations of the material's structure with the electronic band structure. This approach has precedence, being used in prediction of novel thermoelectrics,\cite{JiaXue2022,NaGyoungS2021,ShengYe2020} construction materials,\cite{MASOODCHAUDRY2021101897} solar cell materials,\cite{ChoudharyKamal2019,doi:10.1021/acs.jpclett.1c03526} and more. 

\begin{figure}[H]
    \centering
    \includegraphics[scale=0.30]{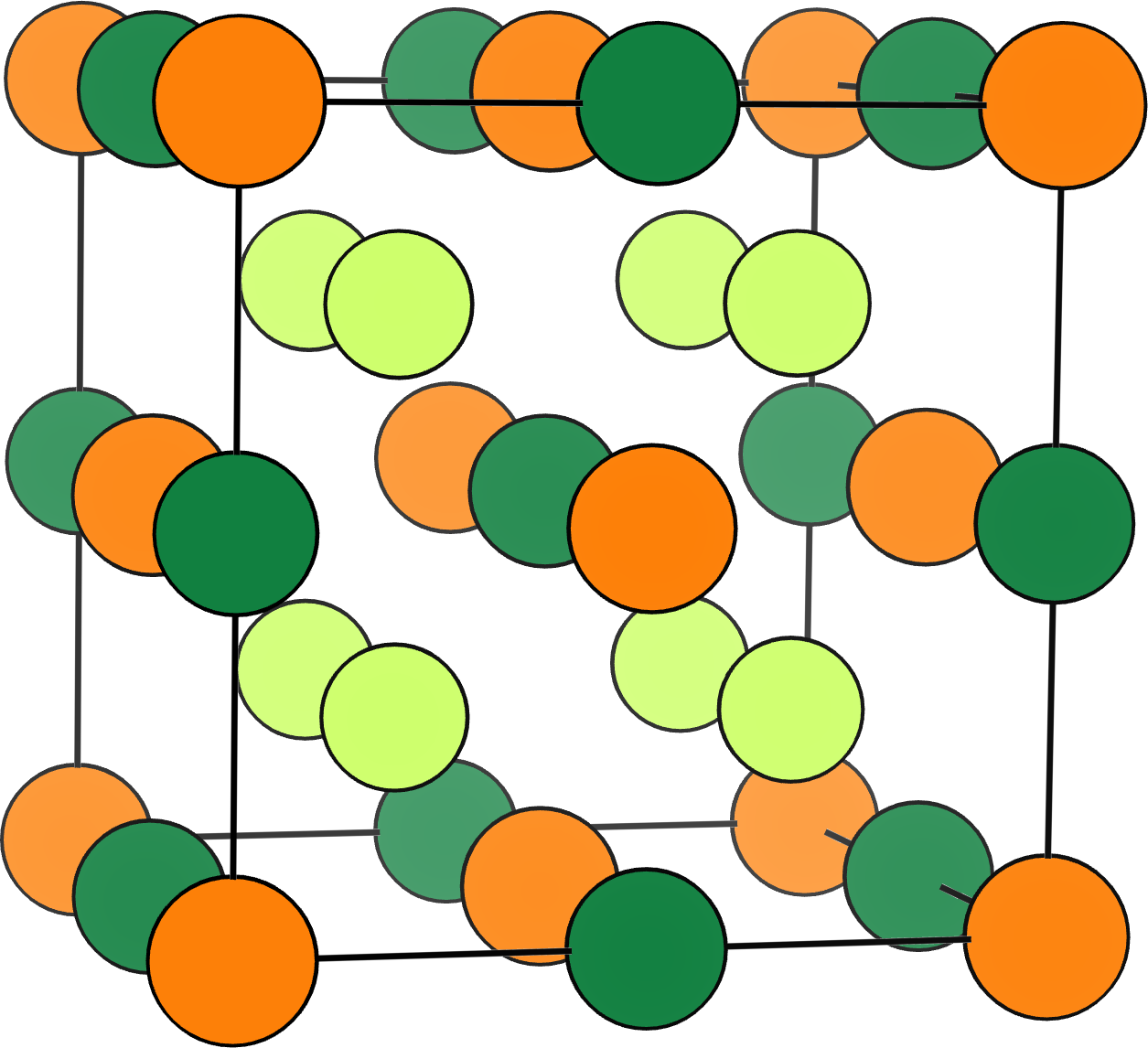}
    \caption{Illustrative full-Heusler crystal---\ch{MnCu2Al} \ch{Al}, \ch{Mn}, and \ch{Cu} are colored in orange, green, and lime respectively. Structure taken from Materials Project\cite{Jain2013} (MP-id: mp-3574). [colors online]
    }
    \label{fig:HeuslersCrystalStructure}
\end{figure}

To begin this approach for QMs, first we narrow our materials of interest to a particularly successful class of QM materials, the Heusler family,\cite{galanakis2016theory,graf2010heusler,casper2012half,graf2011simple,poon2018recent,chibani2018first} which has attracted major interest from the community for their wide-ranging applications in the domain of spintronics,\cite{birsan2020zr,khandy2019full,casper2012half} thermoelectrics,\cite{zaferani2019strategies,poon2018recent,chibani2018first} superconductors,\cite{hu2021charge,xiao2018superconductivity,uzunok2020physical,kautzsch2019aupdtm} solar-cells,\cite{abdel2021high,murtaza2021lead,wang2020photoactive,leoncini2022correlating} multifunctional topological insulators,\cite{sattigeri2021dimensional,barman2018topological,zhang2020machine} etc. Chemically, they are ternary compounds which can be further distinguished into ``half'' and ``full'' Heuslers based on their stoichiometry. The chemical formula for half-Heuslers and full-Heuslers are of the form \ch{XYZ} (1:1:1) and \ch{X2YZ} (2:1:1) respectively, where \ch{X} and \ch{Y} generally have a cationic character (\textit{d}-block elements--transition metals) and Z is expected to be the anionic counterpart (\textit{p}-block elements). Crystallographically, Heusler compounds exhibit a face-centered cubic crystal structure, as depicted on Figure \ref{fig:HeuslersCrystalStructure}, with space groups \num{216} and \num{225} for half-Heuslers and full-Heuslers respectively. As a myriad of chemical choices are possible with the X, Y and Z sites, Heusler's material properties can be tuned by altering the composition, which can be leveraged for accelerating material development in applications.

The chemical composition to EBS dependency for Heuslers is, however, a complicated relationship, being dependent on which atoms are located at which atomic site in the crystal structure and the resulting interaction of their atomic orbitals.\cite{ojala2010permutation,ma2017computational,ma2018computational} 
A challenge in Heusler design is to predict how the chemical composition and crystallographic arrangement influences EBSs without requiring exhaustive DFT calculation. However, given the large number of possibilities, comprising $>$ \num{1500} known members with an exponentially larger number of possibilities if different atomic crystal arrangements are considered for a given composition, first-principles calculation and learning of their influence on the EBS is unmanageable, making it a good test bed for attempting to use AI and ML as mentioned above.
We create a Heuslers dataset using the Materials Project (MP) database;\cite{Jain2013} considering compounds with cubic crystal structures and general formulas \ch{XYZ} and \ch{X2YZ}, which correspond to space groups 216 and 225 (a total \num{276} compounds). The EBSs are also retrieved from the MP which are computed using DFT calculations. Using this Heuslers dataset along with the python library Matminer\cite{WARD201860, Matminer} for rapid features extraction, we developed an ML model that correlates Heusler's chemical compositions with their EBSs; specifically with the number of band crossings within a specific range of the ground state energy ($\pm$\SI{1}{\electronvolt}). We extract the number of crossings from the EBS using an automated algorithm to mine the EBS for that particular pattern. The ML model is then developed using the XGBoost algorithm\cite{chen2016xgboost,chen2015xgboost}---a parallel tree boosting technique under Gradient Boosting framework---to train the model as a function of periodic table properties of the materials constituent elements (chemical composition) and global crystal structure parameters. Also, since the Heuslers dataset consists of a relatively small number of compounds (\num{276}), we also created a dataset of cubic compounds removing the restriction of specific transition metal atoms for \ch{X} and \ch{Y} elements, and \textit{p}-block atoms for \ch{Z} elements, but retaining the space groups \num{216} and \num{225}. This ``Cubic dataset'' consists of \num{3751} structures, and is used to benchmark the ML models and understand the generalizability of the insight gained from the Heuslers dataset.

\section{Methodology}

\subsection{Materials Data-set}
\begin{figure}[H]
    \centering
        \subfigure[]{
        \includegraphics[width=3in]{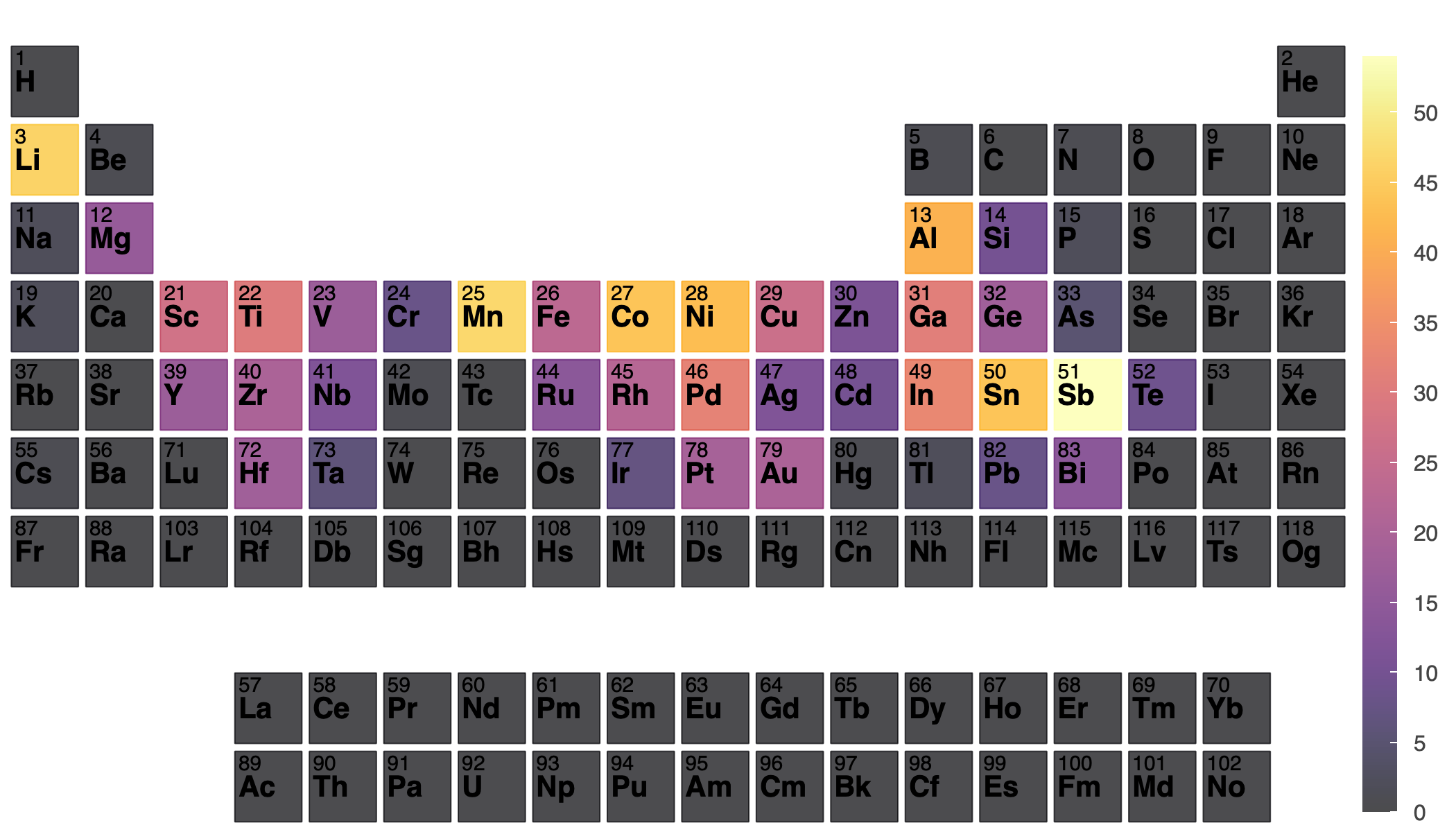}
        \label{fig:small_dataset_compo}
        }
        \quad
        \subfigure[]{
        \includegraphics[width=3in]{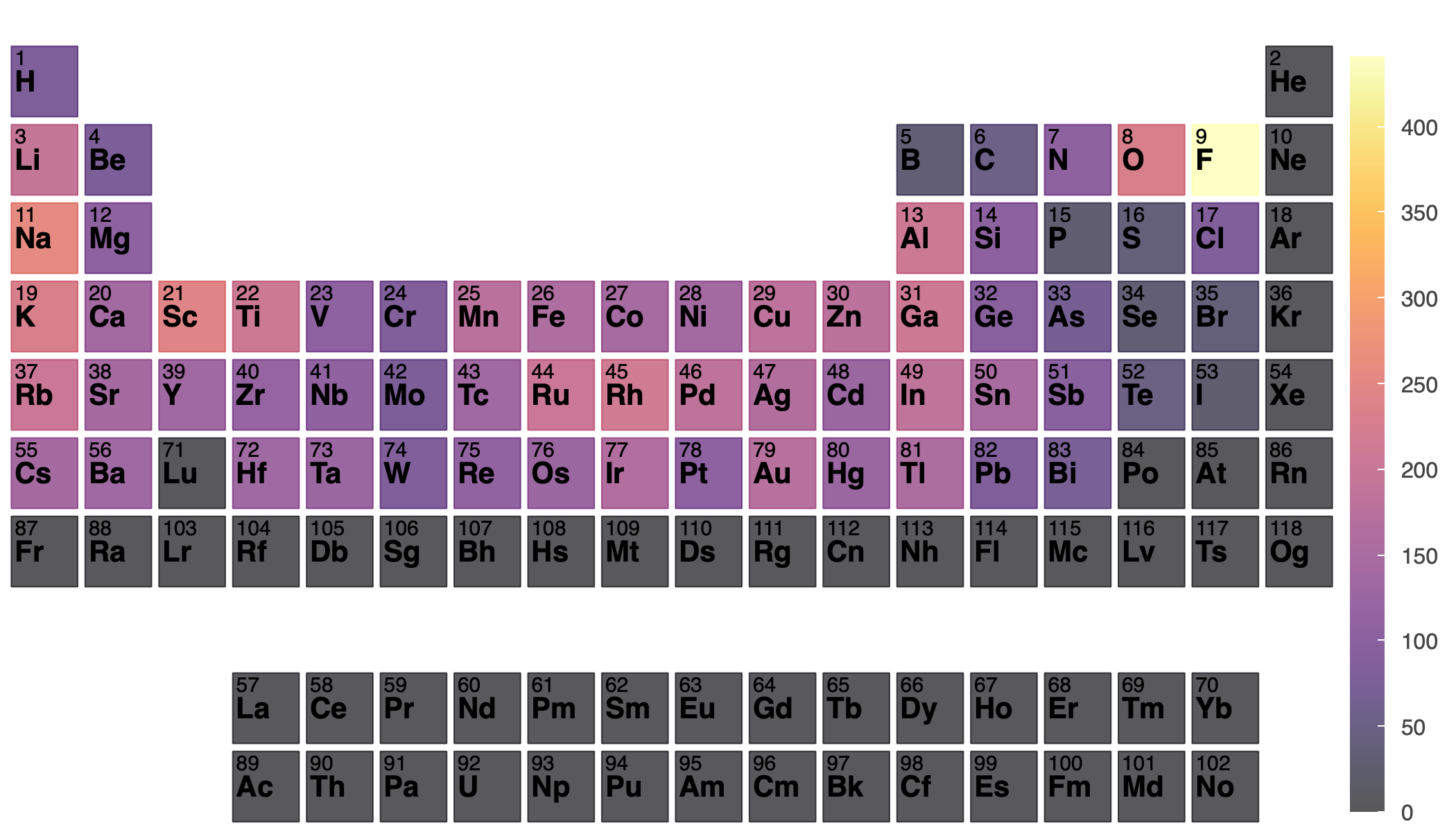}
        \label{fig:big_dataset_compo}
        }
\caption{a) Periodic table heatmap for element appearance on Heuslers dataset, b) Periodic table heatmap for element appearance on the Cubic dataset. Elements are colored according to their appearance on in the respective datasets, gray corresponds to no appearance and yellow for maximum appearance. [Colors online]}
\label{fig:dataset_compo}
\end{figure}

The data used in this work is entirely sourced from the the Materials Project with $\sim$ \num{76240} entries of EBSs calculated via DFT. 
For the Heusler and Cubic dataset with space groups of \numlist{216;225}, the whole MP database was searched---using the pymatgen python library\cite{ONG2013314}---for materials with cubic crystal symmetry (space groups \numlist{216;225}) which were then downloaded including their chemical, structural, EBS. The total number of materials found after this search was \numlist{276;3751} for the Heuslers and Cubic datasets respectively; and are available at \href{https://github.com/Material-Mind/Heuslers-EBS-Dirac-points.git}{GitHub Link}. Heuslers dataset is formed by \numlist{107;169} half (space group 216) and full (space group 225) Heuslers respectively. Consisting of compounds comprised by \ch{X} and \ch{Y} elements from groups \numrange{1}{12} (\textit{s} and \textit{d} blocks) and periods \numrange{4}{6}; and \ch{Z} elements from groups \numrange{13}{16} and (\textit{p}-block) periods \numrange{3}{6} (see Figure \ref{fig:small_dataset_compo}). On the other hand, the Cubic dataset consist of compounds comprised by \ch{X} and \ch{Y} elements from groups \numrange{1}{12} (\textit{s} and \textit{d} blocks) and periods \numrange{1}{6}; and \ch{Z} elements from groups \numrange{13}{17} (\textit{p}-block) and periods \numrange{2}{6} (see Figure \ref{fig:big_dataset_compo}). The Heuslers dataset compounds are constrained to have for \ch{X} and \ch{Y} elements \textit{s} and \textit{d} blocks atoms whereas for \ch{Z} element \textit{p}-block atoms. These constraints are removed for the Cubic dataset compounds i.e. \ch{X}, \ch{Y}, and \ch{Z} can be either \textit{s}, \textit{d} or \textit{p}-block elements. However, the symmetry constraints (space groups 216 or 225) were retained. Noteworthy, \ch{X}, \ch{Y} and \ch{Z} elements in the Cubic dataset exhibit counting ranging from $\sim$\numrange{75}{225}, highlighting fluorine (\ch{F}) with an outlier counting of \num{450}. In contrast, \ch{X}, \ch{Y} and \ch{Z} elements in the Heuslers dataset exhibit counting ranging from $\sim$\numrange{7}{55}. 

\subsection{Electronic Band Structure: Dirac points identification - Number of Crossings}
\begin{figure}[H]
        \subfigure[]{
        \includegraphics[scale=0.29]{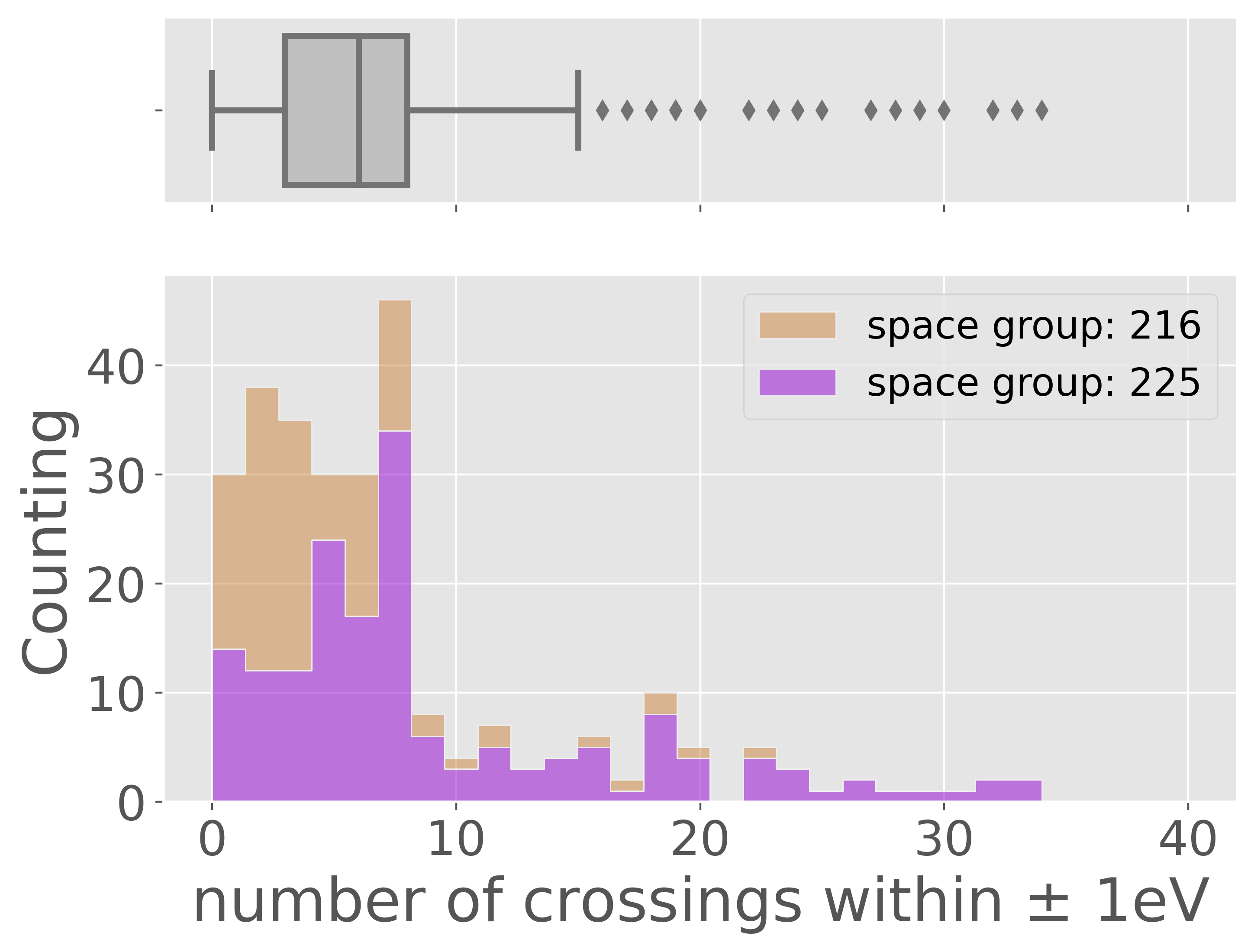}
        \label{fig:small_dataset_croos}
        }
        \quad
        \subfigure[]{
        \includegraphics[scale=0.29]{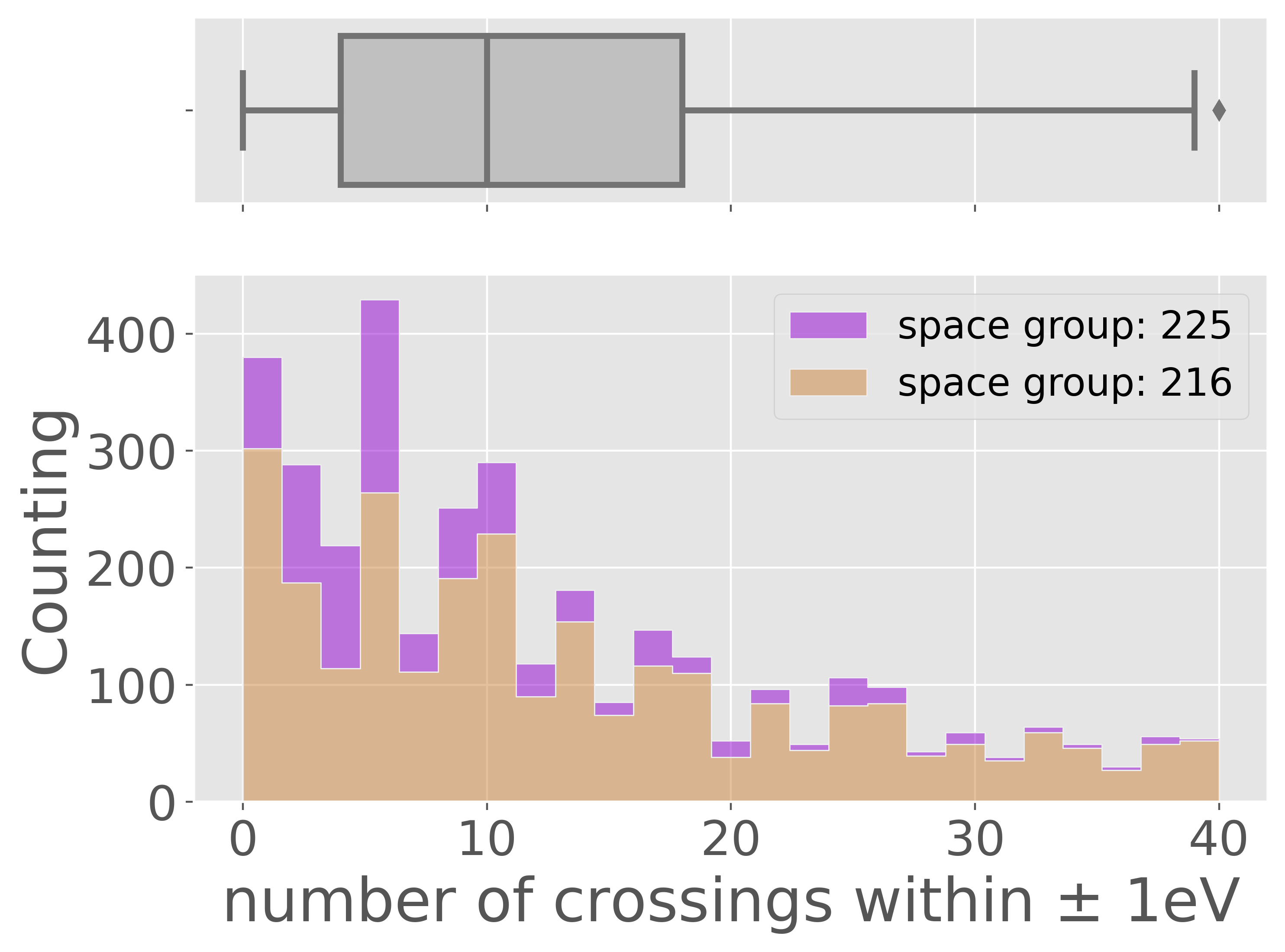}
        \label{fig:big_dataset_croos}
        }
\caption{Distribution of number of crossings on the EBS (within $\pm$\SI{1}{\electronvolt}) a) Heuslers dataset, and b) Cubic dataset. [Colors online]}
\label{fig:dataset_croos}
\end{figure}

EBSs are imported as a $N \cdot K$ matrix of energy value, with N being the number of bands calculated by the DFT calculation and K being the number of momentum-points sampled along the MP-recommended BZ path. The MP database uses the BZ path convention presented in \cite{setyawan2010high}. 
Specifically we identify Dirac points over the EBSs---termed 'number of crossings'---within a specific range of the ground state energy ($\pm$\SI{1}{\electronvolt}). The number of crossings distributions for the Heusler and Cubic datasets are shown in Figure \ref{fig:dataset_croos}. In both cases, the distributions are right skewed but despite the Cubic dataset having \num{14} times more compounds than Heuslers dataset, both number of crossings distributions have close medians $\sim$\num{8}. Suggesting the distributions shape correspond to the number of crossings natural variation instead of an data errors or sampling problems. Also, more than \SI{50}{\percent} of the Heuslers number of crossings range consists of outlier data points (see box plot on Figure \ref{fig:small_dataset_croos}), whereas the Cubic dataset distribution has almost no outliers (see box plot on Figure \ref{fig:big_dataset_croos}).

\subsection{Machine Learning Modeling}
\begin{figure}[H]
    \centering
    \includegraphics[width=0.7\textwidth]{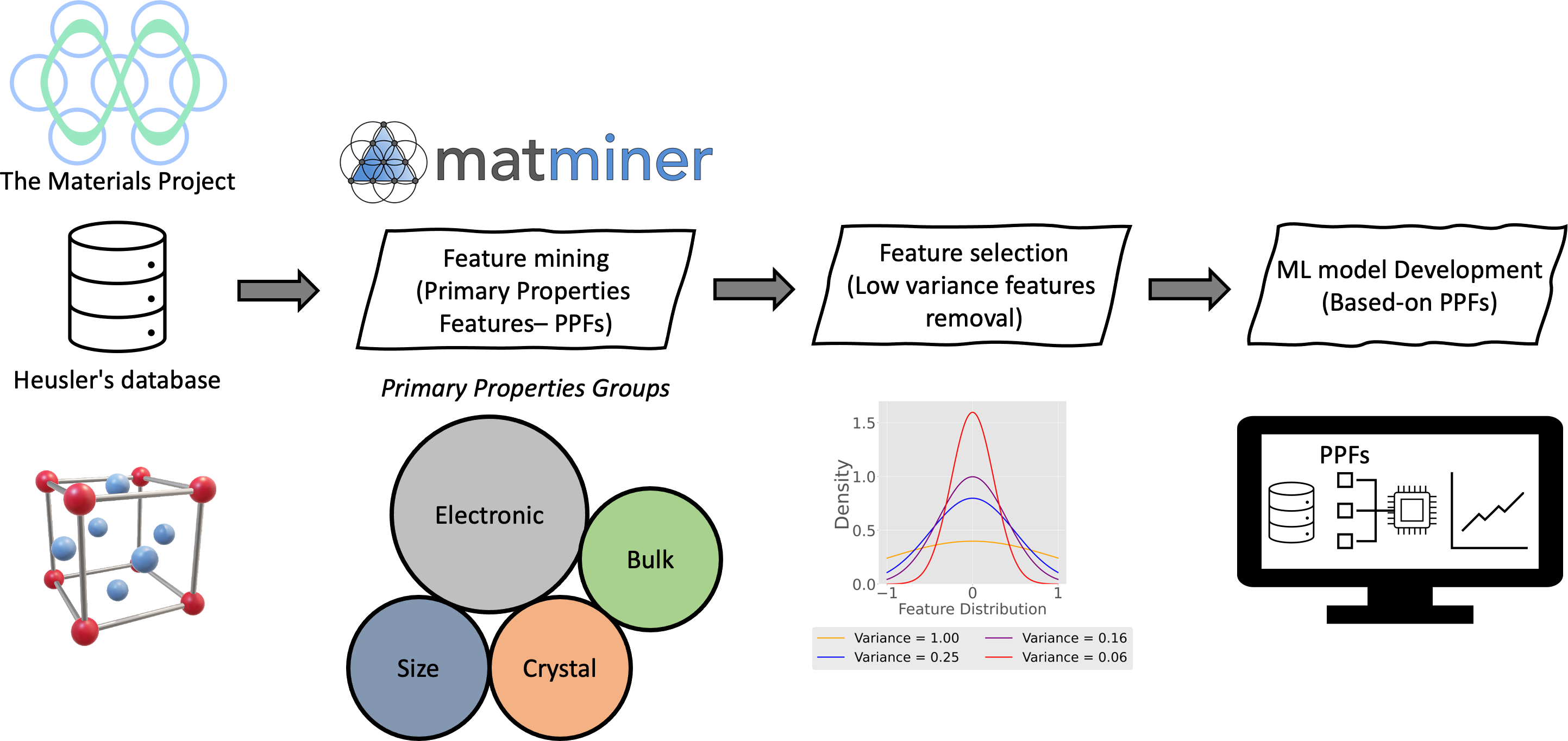}
    \caption{ML modeling flowchart; which consist of: 1) data minig and data sets creation, 2) features extraction, 3) features selection, and 4) ML models development. [Colors online]}
    \label{fig:MLWorkFlow}
\end{figure}

A Machine learning (ML) pipeline is implemented to develop models that can predict the number of crossings in the EBSs (within $\pm$\SI{1}{\electronvolt}) for the Heuslers and Cubic datasets. From the logic that the EBS's number of crossings depends on the nature of the atoms within the crystal lattice and their arrangement with each other, the models take as the input as the identities of the atoms in the crystal---encoded as elemental properties---producing, as output, estimates of the number of crossings (within $\pm$ \SI{1}{\electronvolt}). A general flowchart for ML pipeline is presented on Figure \ref{fig:MLWorkFlow}; specifically, the features engineering consists of two main steps: 1) Feature extraction, and 2) Feature selection. For features extraction, atom properties are taken from the python library Matminer\cite{ward2018matminer,Matminer} for both datasets and named primary properties features (PPFs) and are classified into 4 groups: bulk, electronic, size, and crystal. Matminer creates overall features i.e., for a specify elemental composition of a crystal, and identifies the minimum and maximum values for a given property. We consider a total of 53 PPFs including e.g, electronegativity, covalent radius, volume per atom, etc.  
The PPFs, for developing the ML models, are selected based on the low-variance criterion; i.e. any given PPF which consists of a constant value in the \SI{80}{\percent} or more of its values distribution, or whose variance is $\mathrm{\leq}$ \num{0.16}, are removed. See SI for the full list of PPFs and their values. 





The ML regression models and features importance are performed with the XGBoost approach \cite{chen2015xgboost,chen2016xgboost}. XGBoost provides a parallel tree boosting algorithm \cite{chen2008trada} that contains three elements: 1) a minimizing loss function, 2) a weak learner (regression tree) for making predictions, and 3) an additive model that combines weak learners into strong ones in order to minimize the loss function. In our model, the loss function is the root mean square error (rmse) between the computed and the ML predicted band structure number of crossings (within $\pm$\SI{1}{\electronvolt}). The regression trees weak learners are set by specifying the learning rate, gamma, max\_depth, max\_delta\_step, subsample, colsample\_bytree, colsample\_bylevel, lambda, alpha, gamma, min\_child\_weight, and scale\_pos\_weight parameters. Bayesian optimization and cross validation (shuffle-split (7-folded)) algorithms, as implemented in scikit-optimize python library,\cite{head_tim_2020_4014775} are employed to identify the best combinations of hyperparameter values (e.g., max depth, number of threads, etc.) during the ML-models development. The Feature importance score (using gain scoring) is estimated to identify those PPFs that most influence the prediction of band structure number of crossings (within $\pm$\SI{1}{\electronvolt}). 
\SI{70}{\percent} of the dataset is assigned to the training set, whereas the remaining \SI{30}{\percent} is assigned to the testing set. The training and test sets are assessed based on the mean absolute errors (MAEs) between the calculated (with DFT) and the predicted (with ML) as illustrated in eq. \ref{eq:MAE}:

\begin{equation} \label{eq:MAE}
\mathrm{
MAE = \frac{\sum_i^N \left|n_i^{DFT} - n_i^{ML}\right|}{N}
}
\end{equation}

\noindent where $\mathrm{n}$ stands for band structure number of crossing (within $\pm$\SI{1}{\electronvolt}).

\section{Results}
\subsection{Machine Learning Model}
\begin{figure}[H]
    \centering
        \subfigure[]
        {
        \includegraphics[width=3in]{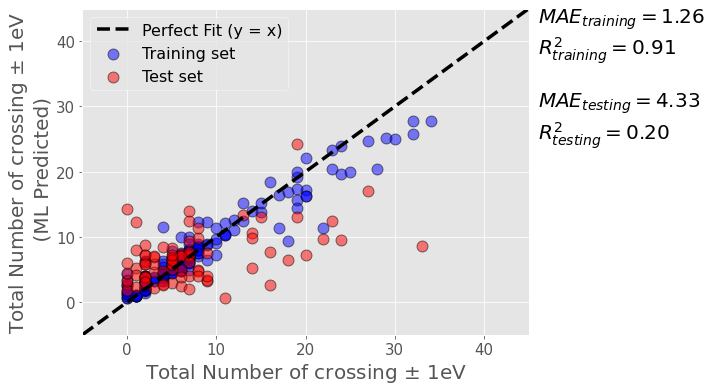}  
        \label{fig:MLmodela}
        }
        \quad
        
        \subfigure[]{
        \includegraphics[width=3in]{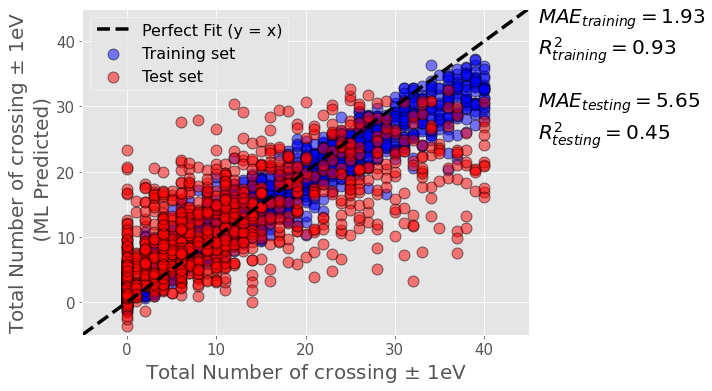}
        \label{fig:MLmodelb}
        }
\caption{ML model parity plot comparing DFT-calculated with ML predicted values for band structure number of crossing (within $\pm$\SI{1}{\electronvolt}) a) Heuslers dataset, and b) Cubic dataset. [Colors online]}
\label{fig:MLmodel}
\end{figure}

Figures \ref{fig:MLmodela} and \ref{fig:MLmodelb} compare, in parity plots, the ML predictions with DFT-calculated EBS's number of crossings (within $\pm$\SI{1}{\electronvolt}) for the Heuslers and Cubic datasets respectively. For these models, the coefficient of determination ($\mathrm{R^2}$) of training and testing sets are \num{0.91} and \num{0.20}, and \num{0.93} and \num{0.45} for the Heuslers and Cubic datasets respectively. Similarly, the mean absolute errors (MAEs) are \num{1.26} and \num{4.43}, and \num{1.93} and \num{5.65}. Comparing the training and testing set's MAEs with the respective number of crossings distribution, the MAEs correspond to the \SI{6.74}{\percent} and \SI{28.9}{\percent}, and \SI{9.50}{\percent} and \SI{20.1}{\percent} of distribution values, for the Heuslers and Cubic datasets respectively. First, the ($\mathrm{R^2}$) of the testing set for the Cubic dataset has improved by nearly a factor of two from the Heuslar dataset, despite the training statistics being nearly the same. This implies that the Heuslar dataset size, which was only about 1/14 the size of the cubic dataset, was a limiting factor. However the MAE of the Cubic dataset's test set is very similar to the MAE of the Heuslar dataset's test set suggesting that the ML performance observed for the Heuslers compounds is not only consequence of the dataset size. And while the Cubic dataset predictions do show a clear linear tendency, the accuracy of the predictions leave much to be desired. Since the ML models features lack of crystal site specific information---which can better encode chemical composition, atoms arrangement, and chemical environment effects, as DFT calculations do---their prediction performance is affected impacting the ML models variance. 

\begin{figure}[H]
    \centering
        \subfigure[]
        {
        \includegraphics[width=3in]{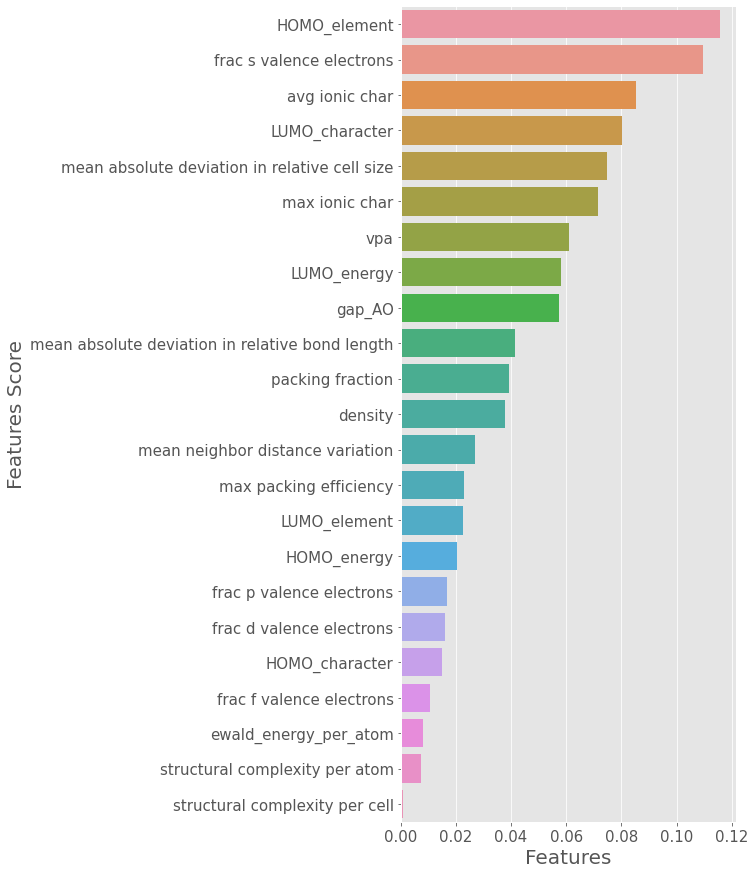}  
        \label{fig:MLmodel_Heuslers_FI}
        }
        \quad
        \subfigure[]
        {
        \includegraphics[width=3in]{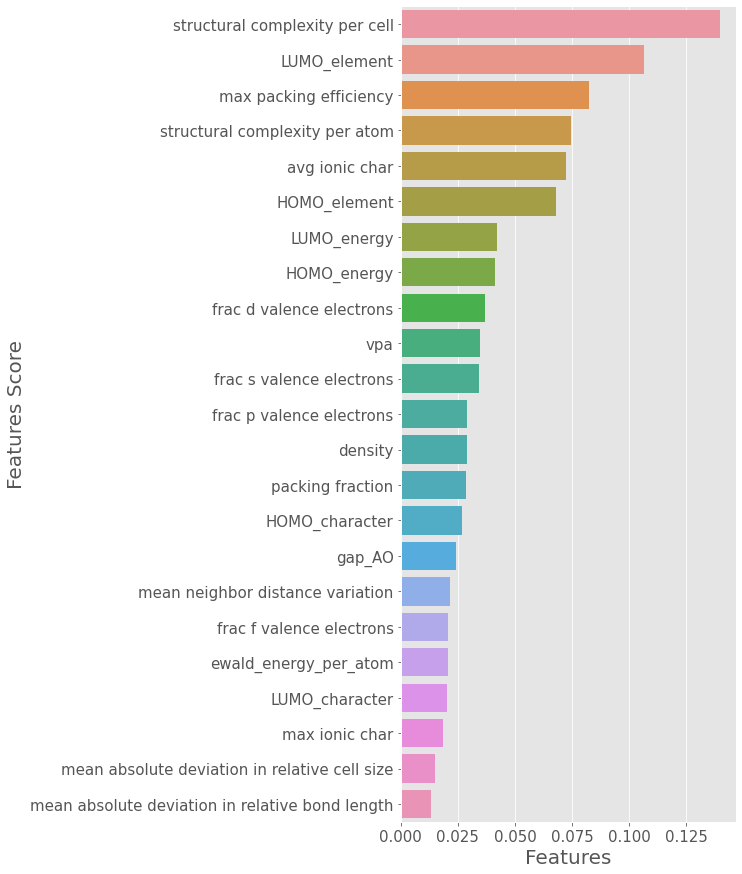}  
        \label{fig:MLmodel_Cubic_FI}
        }
\caption{a) Feature important scoring (Gain) for a) Heuslers compounds ML model, and b) Cubic compounds ML model. [Colors online]}
\label{fig:MLmodelFI}
\end{figure}

Regarding the ML models features importance, Figure \ref{fig:MLmodelFI} shows the bar plots for the features important score (gain). 
Highlighting for both ML models---among the first 10 most important features---the ``HOMO element'', ``avg. ionic char'', ``vpa'' (volume per atom), and ``LUMO energy''. All of these features are associated with both electronic and crystal primary properties. Moreover, by grouping features importance scores---based on the electronic and crystal categories---the electronic properties have an slightly higher importance in predicting the number of crossing on the ML models, as expected from comparison to DFT. Among the electronic features, the ones related to HOMO and LUMO frontier levels appear in both models; similarly, for crystal features, the ``structural complexity per cell'', a composite feature related to the local coordination of the atoms, is ranked highly, again as expected from DFT. In the case of bulk properties like density, they do not have a significant importance among the ML models features, as expected from DFT. Taken together, these results imply the model did begin to capture the correct underlaying physics driving the target result, but was hampered by the lack of site-specific features. Among the global features it had access to, it ranked most highly the ones which were at least partially site-specific; structural complexity, atomic orbitals, etc. 

\section{Conclusions}
In this work, we used machine learning to develop a model correlating composition vs. number of Dirac points (called the number of crossings) in the electronic band structures for Heuslers and other Cubic compounds by identifying said crossings using an automated algorithm as well as generating chemical composition and global crystal structure features. In general, our ML model captured the overall trend in the Heuslers, predicting the EBS number of crossings; however, the ML model (parity plot) suffered from  significant variance. The Heuslers datasets size (\num{276} compounds) was not the limiting factor in regards of the variance; an additional ML model was created for the larger dataset of Cubic compounds of size ($\sim$\num{3800} compounds), and it also exhibited a similar variance. This is, however, within expectation, due to the nature of the EBS where it is well understood that atomic site specific properties determine the band structure. A methodology for handling atomic site specific features has to be developed such that ML models can incorporate their description, in a better match to the underlying quantum mechanics governing the properties, and capture the electronic properties in a more generalized approach. This atomic site specific approach is, however, difficult due to the feature dimensionality dramatically increasing as, for example, the number of possible atomic sites (up to 64) convolved with the number of possible elements at each site (up to 118) convolved with the number of atomic orbitals at each site (up to 14), becomes unuseably large. Thus, an open challenge for future applications of ML in electronic materials intelligence, requires the development of featurizers that are able to generate atomic site features but also encoding them into artificial structural features to reduce the dimensionality.



\bibliography{mybibfile}

\end{document}